\documentclass[12pt]{iopart}
\usepackage{graphicx}

\def\ket#1{\mathinner{|{#1}\rangle}}

\def\imagi{\mathrm{i}}
\def\eulere{\mathrm{e}}
\def\diff{\mathrm{d}}

\def\reff#1{(\ref{#1})}

{\catcode`\|=\active
\gdef\Braket#1{\left<\mathcode`\|"8000\let|\BraVert {#1}\right>}}
\def\BraVert{\egroup\,\mid\,\bgroup}

\begin{document}

\title{Harmonium in intense laser fields: excitation, absorption, and transparency}

\author{O.\ Kidun and D.\ Bauer}
\address{Max-Planck-Institut f\"ur Kernphysik, Postfach 10 39 80, 69029 Heidelberg, Germany }

\begin{abstract}
It is known that the dynamics of two (Coulomb-interacting)
nonrelativistic electrons confined by a parabolic potential and
driven by a classical, intense laser field (in dipole approximation)
is exactly soluble. We calculate the time-dependent population of
the harmonic oscillator states and the energy absorbed from the
laser. It turns out that the key entity on which all observables
sensitively depends is the modulus square of the Fourier-transformed
vector potential of the laser field, evaluated at the harmonic
oscillator frequency. The system is transparent to laser field
configurations for which this entity vanishes. We discuss the
Poisson statistics behavior of the transition probabilities and
analyze the conditions for the complete survival and full depletion
of the initial state.
\end{abstract}

\section{Introduction}
The parabolic well is a model of great practical importance, as it
approximates any arbitrary potential close to equilibrium. In
nanotechnology, potentials of simple shape such as quantum dots are
often well approximated by parabolic potentials. Superpositions of
harmonic oscillators are also used to describe continua,
environments, and field modes.

Correlations, for instance, those introduced by the Coulomb
interaction between electrons, attract growing attention. Correlated
electron dynamics can nowadays be analyzed experimentally with great
precision using ``reaction microscopes''\cite{Ullrich}. Two
electrons confined in a parabolic potential, sometimes called
``harmonium'' or the ``Hooke's atom'' version of
helium~\cite{Kestner}, is of great interest since the problem
represents one of the few exactly soluble problems involving
correlation~\cite{Taut,Kais}. As such harmonium serves as a
testing ground for, e.g.\ new exchange-correlation potentials in
density functional theory.

In the field of strong laser atom interaction, exactly soluble
models are rare as well, in particular as it comes to electron
correlation. Fortunately, the two-electron wavefunction of the
laser-driven harmonium can still be expressed in analytical form as
long as the laser field is treated classically and in dipole
approximation~\cite{Schwengelbeck}.

The outline of the paper is as follows. In Sec.~\ref{sec2} we derive the exact solution of the time-dependent Schr\"odinger equation of two interacting electrons confined in a parabolic potential and driven by a classical (laser) field (in dipole approximation). In Sec.~\ref{sec3} the probabilities for two-electron excitations from an arbitrary state $m$ to a state $n$ are calculated. The latter are the basis for the discussion of optimal energy absorption and transparency in Sec.~\ref{Sec.EA}. Finally, we conclude in Sec.~\ref{sec5}.


\section{Exact states of two-electrons in a parabolic well driven by a laser field}\label{sec2}
In the absence of the laser field the Hamiltonian of harmonium
 reads\footnote{Spin is omitted for brevity.}
\begin{eqnarray}\label{totalSHr1r2}
H({\bf r}_1,{\bf r}_2)=-\frac{\hbar^2}{2m}\nabla_{{\bf
r}_1}^2-\frac{\hbar^2}{2m}\nabla_{{\bf
r}_2}^2+\frac{m}{2}\omega^2{\bf r}^2_1+\frac{m}{2}\omega^2{\bf
r}^2_2+\frac{e^2}{\left|{\bf r}_1-{\bf r}_2\right|} \label{Hamil}
\end{eqnarray}
with $\omega$ the harmonic oscillator frequency and $m$ the mass of
the electron. The Hamiltonian \reff{Hamil} is separable when written
in the center-of-mass (CM) and relative coordinates ${\bf
R}=\frac{1}{2}({\bf r}_1+{\bf r}_2)$ and ${\bf r}={\bf r}_1-{\bf
r}_2$, respectively:
\begin{eqnarray}\label{totalSHRr}
H=H_{R}+H_{r}=\left(-\frac{\hbar^2}{4m}\nabla_{\bf
R}^2+m\omega^2{\bf R}^2\right)+\left(-\frac{\hbar^2}{m}\nabla_{\bf
r}^2+\frac{m}{4}\omega^2{\bf r}^2+\frac{e^2}{\left|{\bf
r}\right|}\right).
\end{eqnarray}
As a consequence, the eigenfunctions $\Phi({\bf r}_1,{\bf r}_2)$  are products of the
eigenfunctions of the CM and relative Hamiltonians $H_R$ and $H_r$,
i.e.\ with $H_R\ket{\phi}=E_R\ket{\phi}$ and
$H_r\ket{\xi}=E_r\ket{\xi}$ follows
\begin{eqnarray}\label{totalSWF}
\Phi({\bf r}_1,{\bf r}_2)\eulere^{-\frac{\imagi}{\hbar}
Et}=\Phi({\bf R},{\bf r})\eulere^{-\frac{\imagi}{\hbar}
(E_R+E_r)t}=\phi({\bf R})\eulere^{-\frac{\imagi}{\hbar} E_Rt} \cdot
\xi({\bf r})\eulere^{-\frac{\imagi}{\hbar} E_rt}.
\end{eqnarray}
Due to the spherical symmetry of both $H_R$ and $H_r$ the
eigenfunctions are conveniently expressed in the form
$\phi_{NLM}({\bf R})=\frac{U_{NL}(R)}{R}Y_{LM}(\Theta,\chi)$ and
$\xi_{n\ell m}({\bf r})=\frac{u_{n\ell}(r)}{r}Y_{\ell
m}(\theta,\varphi)$.

The CM radial wavefunction and energy are readily obtained from the
standard problem of the 3D harmonic oscillator to give
$U_{NL}(R)=C_{NL} R^{L+1}H_N(R)\eulere^{-m\omega R^2/\hbar}$ and
$E_R\!\!=\!\!\hbar\omega(2N+L+\frac{3}{2})$.
$H_N\!\!=\!\!\!_1F_1(-N,L\!+\!\frac{3}{2},m\omega R^2/\hbar)$ is the
$N$-th order Hermite polynomial, $C_{NL}$ is the normalization
constant. The radial wavefunction $u_{n\ell}(r)$ has the closed
analytical form (unnormalized)
$u_{n\ell}(r)\!\!=\!\!r^{\ell+1}\eulere^{-m\omega
r^2/4\hbar}\sum_{k=0}^{n-1}A^kr^k$ only for certain oscillator
frequencies \cite{Taut}. For other frequencies, it has to be
evaluated numerically. Once it is found, the eigenenergy has the
simple form $E_r=\hbar\omega(n+\ell+\frac{1}{2})$.

Adding the interaction with an electromagnetic field affects only
the CM Hamiltonian so that the problem still separates. Having in
mind the interaction with near infrared or visible laser pulses, we
adopt the dipole approximation and obtain
\begin{equation}\label{totalTDH}
\fl\qquad H=\left(\!-\frac{\hbar^2}{m}\nabla_{\bf r}^2+\frac{m}{4}\omega^2{\bf
r}^2\!+\frac{e^2}{\left|{\bf
r}\right|}\right)\!+\!\left(\!\frac{1}{2\cdot2m}\left[-\imagi\hbar\nabla_{\bf
R}\!+\!\frac{2e}{c}{\bf A}(t)\right]^2\!\!\!+m\omega^2{\bf
R}^2\!\right)
\end{equation}
with ${\bf A}(t)$ the vector potential and  $c$ the light velocity.
The electric field of the laser is given by ${\bf E} =\! -\partial_t
{\bf A}$ while the magnetic field is neglected\footnote{Two
parabolically confined electrons in constant magnetic fields were
studied in \cite{Taut94}.} in dipole approximation. The total wave
function reads
\begin{eqnarray}\label{totalTDWF}
\Psi({\bf R},{\bf r},t)=\psi({\bf R},t)\cdot\xi({\bf
r})\eulere^{-\frac{\imagi}{\hbar} E_rt}
\end{eqnarray}
with $\xi({\bf r})$ the same as before.

The solution of the time-dependent  Schr\"odinger equation governing
the CM motion with twice the electron mass $\mu=2m$ and charge
$\epsilon=2e$
\begin{eqnarray}\label{cmtd}
\left(-\imagi\hbar\partial_t+\frac{1}{2\mu}
\left[-\imagi\hbar\nabla_{\bf R}+\frac{\epsilon}{c}{\bf
A}(t)\right]^2+\frac{1}{2}\,\mu\omega^2{\bf R}^2\right)\psi({\bf
R},t)=0
\end{eqnarray}
is known \cite{Husimi,Uryupin}. In the case of a linearly polarized
laser field ${\bf A}(t)=A(t){\bf e}_z$ one has
\begin{eqnarray}\label{CMTDWF}
\psi({\bf R},t)=U(X)\eulere^{-\frac{\imagi}{\hbar}
E_Xt}U(Y)\eulere^{-\frac{\imagi}{\hbar} E_Yt}U(Z,t),
\end{eqnarray}
with the two unaffected harmonic oscillator eigenfunctions
$U(X)=C_{N_X}H_{N_X}\!(X)\eulere^{-\mu\omega X^2/2\hbar}$,
$U(Y)\!=\!C_{N_Y}H_{N_Y}(X)\eulere^{-\mu\omega Y^2/2\hbar}$ and the
respective eigenenergies  $E_{X}=\hbar\omega(N_{X}+\frac{1}{2})$,
$E_{Y}\!\!=\!\hbar\omega(N_{Y}+\frac{1}{2})$.  Instead, the
eigenfunction $U_{N_Z}(Z)$ becomes dressed by the laser field and reads
\begin{eqnarray}\label{ZCMTDWF}
U_{N_Z}(Z,t)=C_{N_Z}H_{N_Z}
\left(\sqrt{\frac{\mu\omega}{\hbar}}\left[Z-\frac{\epsilon g(t)}{\mu
c}\right]\right)
\times\hspace{40mm}\nonumber\\
\times \exp\left\{-\imagi\omega\!\left(N_Z+\frac{1}{2}\right)\!t+
\frac{\imagi \epsilon^2\omega^2}{2\mu\hbar c^2}\int_{t_0}^t
\left(f^2(\tau)\!-\!g^2(\tau)\!-\!\frac{A^2(\tau)}{\omega^2}\right)\diff\tau+
\right.\nonumber\\
\left.+\frac{\imagi \epsilon\omega}{\hbar c}
f(t)\left[Z-\frac{\epsilon g(t)}{\mu c}\right]
-\frac{\mu\omega}{2\hbar}\left[Z-\frac{\epsilon g(t)}{\mu
c}\right]^2 \right\}.
\end{eqnarray}
The normalization constants are given by
$C_k=\left(\!\sqrt{\pi\hbar/\mu\omega}\,\,\,2^{k}k!\right)^{-1/2}$.
The laser field is turned on at time  $t=t_0$. Before, the system is
assumed to be in an eigenstate determined by the quantum numbers
$N_X$, $N_Y$, and $N_Z$. The functions $f(t)$ and $g(t)$ are given
by
\begin{eqnarray}
f(t)=\cos\omega t\int_{t_0}^t A(\tau)\sin\omega\tau \diff\tau
-\sin\omega t\int_{t_0}^t A(\tau)\cos\omega\tau \diff\tau \ , \label{ft}\\
g(t)=\sin\omega t\int_{t_0}^t A(\tau)\sin\omega\tau \diff\tau
+\cos\omega t\int_{t_0}^t A(\tau)\cos\omega\tau \diff\tau \ .\label{gt}
\end{eqnarray}

>From the structure of (\ref{ZCMTDWF}) one can infer {\em (i)} that
the center of the CM wave packet describes motion along a trajectory
${\bf R}(t)=[0,0,\frac{\epsilon}{\mu c}\,g(t)]$ where, in fact,
$g(t)$ is proportional to the excursion of a driven, classical
harmonic oscillator, and {\em (ii)} that the time-dependent solution
$U_0(Z,t)$ represents a so-called {\it coherent} wave packet, i.e.\
a state of minimum uncertainty $\Delta p\ \Delta Z=\hbar/2$, equally
distributed over the momentum $p\ $ and the spatial coordinate $Z$
(see, e.g.\ \cite{Scully}).

\section{Photoexcitation of the electron pair} \label{sec3}
Let us consider the photoexcitation of an electron pair confined in
a parabolic well (e.g.\ two electrons occupying the low-lying states
of a quantum dot). The transition amplitude is given by the overlap
of the exact time-dependent two-electron wave function $\Psi({\bf
r}_1,{\bf r}_2,t)$ with the asymptotic stationary solution
$\Phi({\bf r}_1,{\bf r}_2)$ when the field is switched off:
\begin{eqnarray}\label{T_gen}
T(t)= \Braket{\Phi\left|\right.\!\Psi(t)}=
\Braket{\phi_{NLM}\left|\right.\psi_{N_xN_yN_z}\!(t)}=
\Braket{U_{K_z}\left|\right.U_{N_z}\!(t)}.
\end{eqnarray}
Here we made use of the fact that neither the sub-problem of
relative motion of the two electrons nor the $X$ and $Y$ components
of the CM motion are affected by the laser field. What is left is
the probability to find the CM quasiparticle occupying a stationary
state $U_{N^{{\rm fin}}_{z}}(Z)$ of the free oscillator after the
action of the laser.

Let us first discuss the case where we start from the ground state, i.e.\ $N_X=N_Y=N^{^{\rm
ini}}_Z=0$, $\langle {\bf R} | \psi(t)\rangle=\psi_{000}({\bf
R},t)$. The squared modulus of the corresponding transition
amplitude to some final state with the quantum number $N_{Z}^{^{\rm
fin}}\equiv n$ reads
{\small
\begin{eqnarray}
\fl\qquad\left|T^{(0\to n)}(t)\right|^2=C^2_0C^2_n\left|\exp\left\{\imagi
n\omega t+\imagi \frac{\epsilon^2\omega^2}{2\mu\hbar c^2}
\int_{t_0}^{t}\left(f^2(\tau)-g^2(\tau)
-\frac{A^2(\tau)}{\omega^2}\right)\diff\tau\right\}\right|^2\times\nonumber\\
\fl\qquad\qquad\times \left|\int_{\!\!-\infty}^\infty\hspace{-4mm}\diff Z\,
H_n\!\!\left(\!\!\sqrt{\frac{\mu\omega}{\hbar}}Z\!\!\right)
\exp\!\!\left\{\!-\frac{\mu\omega}{2\hbar}
Z^2\!+\!\imagi\frac{\epsilon\omega}{\hbar c}
f(t)\!\left[\!Z\!-\!\frac{\epsilon g(t)}{\mu c}\!\right]\!-\!
\frac{\mu\omega}{2\hbar}\!\left[\!Z\!-\!\frac{\epsilon g(t)}{\mu
c}\!\right]^{2}\! \right\}\!\right|^2\!\!.
\end{eqnarray}
}%
One can omit all purely time-dependent imaginary exponents,
producing unity, and reduce the integral to the table
form~\cite{Grad1} by change of variables
\begin{eqnarray}
\left|T(t)\right|^2\!=
C^2_0C^2_n\eulere^{-\frac{\epsilon^2\omega}{2\mu\hbar c^2}
(g^2(t)+f^2(t))}\!\!
\left|\int_{\!\!-\infty}^\infty\hspace{-1mm}\diff x\,H_n(x)\eulere^{-(x-y)^2}
\right|^2
\end{eqnarray}
with $x=\sqrt{\frac{\mu\omega}{\hbar}}Z$ and
$y=\frac{\epsilon}{2c}\sqrt{\frac{\omega}{\mu\hbar}}\,\left(g(t)+\imagi
f(t)\right)$.

As a result, the probabilities to find the CM of the electron pair
in the ground (that is the survival probability) or in the $n$-th
excited state are
\begin{eqnarray}
\label{SurvProb} P^{(0\to 0)}(t)&=&\left|T^{(0\to 0)}(t)\right|^2=
\eulere^{-\sigma},
\\\label{ExcProb}
P^{(0\to n)}(t)&=&\left|T^{(0\to k)}(t)\right|^2=
\frac{\sigma^n}{n!}\,\, \eulere^{-\sigma},
\\
\sigma(t)&=&\frac{e^2\omega}{2\mu\hbar
c^2}\,\left[g^2(t)+f^2(t)\right].
\end{eqnarray}
where we introduce the important parameter $\sigma(t)$, which
depends on the laser field and the oscillator frequency [cf.\
eqs.~\reff{ft} and \reff{gt}].

The formula for the photoexcitation probability (\ref{ExcProb})
describes the distribution of the electronic wave packet among the
oscillator states as a function of time and the EM field parameter.
It has the form of a Poisson distribution, which is characteristic
for stochastic processes where the system absorbs the first amount
of energy from the field independently from the possibility of the
absorption of the second portion and so on.

\begin{figure}[htp]
\includegraphics[scale=0.1]{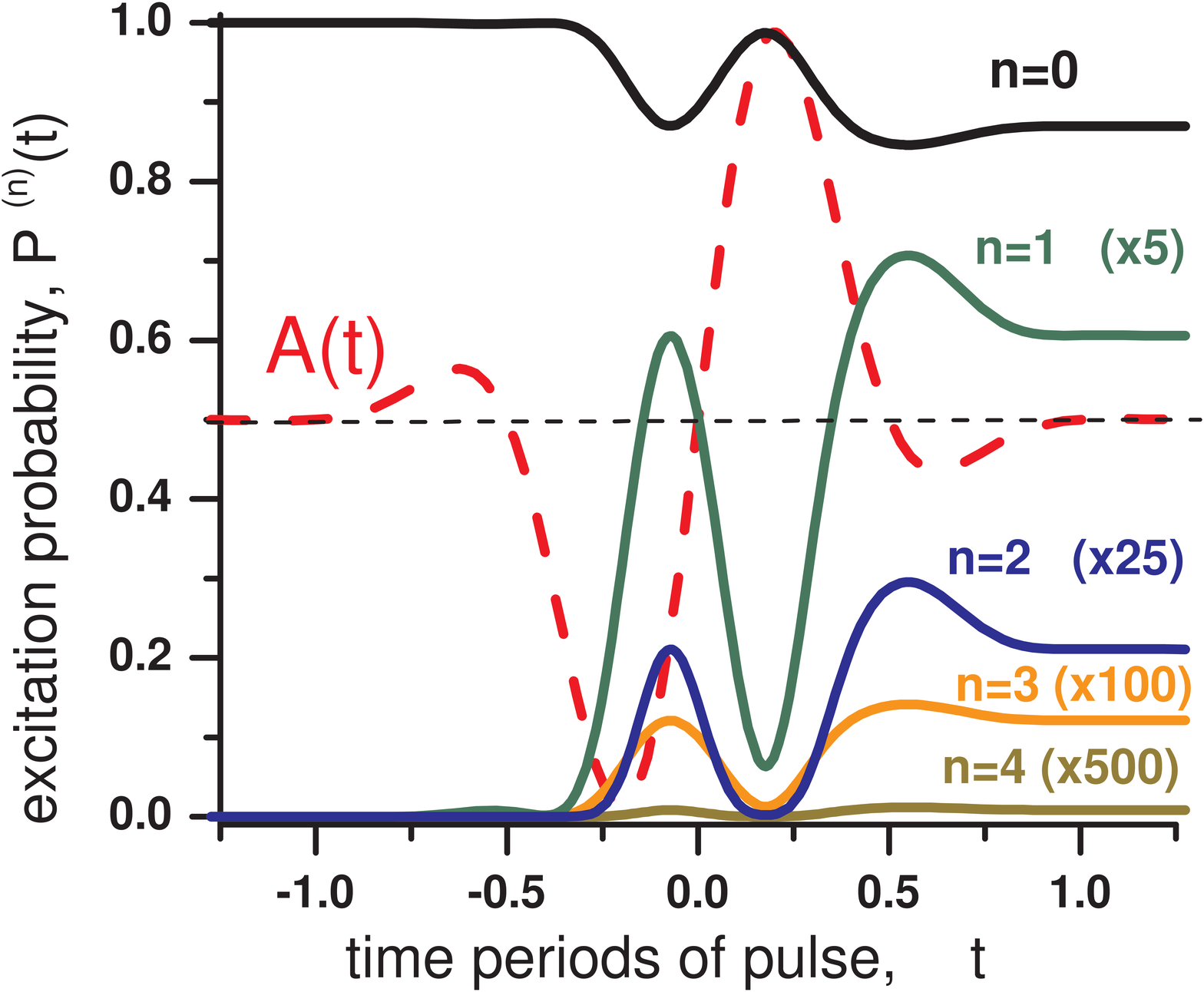}
\caption{The time evolution of the survival probability in the
ground state, $n=0$ (black curve), and the excitation probabilities
to the states with the quantum numbers $n=1,2,3,4$ (green, blue,
orange and olive curves, respectively) is shown together with the
laser pulse (red dashed line)
$A(t)=\sin\omega_lt\,\eulere^{-\left(t/t_p\right)^2/2}$. The laser
frequency, the half-width of the pulse and the oscillator frequency
were chosen $\omega_l=2$\,au, $t_p=1$\,au, and $\omega=1$\,au
respectively. For better visibility, the fast decrease of the
probabilities with the quantum number $P^{(0\to n)}\sim
\frac{1}{n!}$ is compensated by multiplication with the factors
given in brackets.}
\label{fig.Pt}
\end{figure}

In Fig.~\ref{fig.Pt} we illustrate the above expression for
$P^{(0\to n)}$ for a laser pulse of frequency $\omega_l = 2$~atomic
units (au) with a Gaussian envelope and halfwidth $t_p=1$\,au,
$A(t)\!=\!\sin\omega_l t\exp\!\left(-\frac{1}{2}(t/t_p)^2\right)$
(dashed curve). The survival probability $P^{0\to 0}(t)$, changing
from unity to some final value, has one maximum during the pulse due
to the oscillation of the CM wave packet inside the potential and
the associated maximum overlap with the ground state as the center
of the wave packet passes through $Z=0$. The transition
probabilities of the excited states behave in the opposite way,
i.e.\ they are minimum at these times.

In the same way as for the ground state excitations [cf.
eqs.~(\ref{T_gen} -- \ref{ExcProb})], the general expression for the
transition probability from an arbitrary initial $m^{\rm th}$
oscillator state ($t\le t_0$) to the final $n^{\rm th}$ state can be
found; it reads
\begin{eqnarray}\label{Pmn}
\hspace{10mm} P^{(m\to n)}(t)=\,\,
{_2F_0}^2\!\left(-m,-n;-\frac{1}{\sigma}\right)\,
\frac{\sigma^{n+m}}{n!\, m!}\,\, \eulere^{-\sigma},
\end{eqnarray}
with the hypergeometric function
\begin{eqnarray}\label{Lmn}
_2F_0\!\left(-m,-n;-\frac{1}{\sigma}\right)=\sum_{k=0}^{\min(n,m)}\hspace{-3mm}
\frac{n!}{(n-k)!}\,\,\frac{m!}{(m-k)!}\,\,
\frac{1}{k!}\,\left(-\frac{1}{\sigma}\right)^k.
\end{eqnarray}

Equation~(\ref{Pmn}) circumscribes the time dependent population of the
two-electron excited states following the laser pulse. As the
Poisson distribution (\ref{ExcProb}) characterizes the stochastic
absorption of $n$ photons, the distribution (\ref{Pmn}) is
responsible for two processes, namely the independent emission of $\,m-k$ photons and the
(re)absorption of $\,n-k$ photons. The polynomial sum (\ref{Lmn}) in
the expression accounts for the transitions through all the allowed
intermediate states $k$.

\begin{figure}
\includegraphics[scale=0.5]{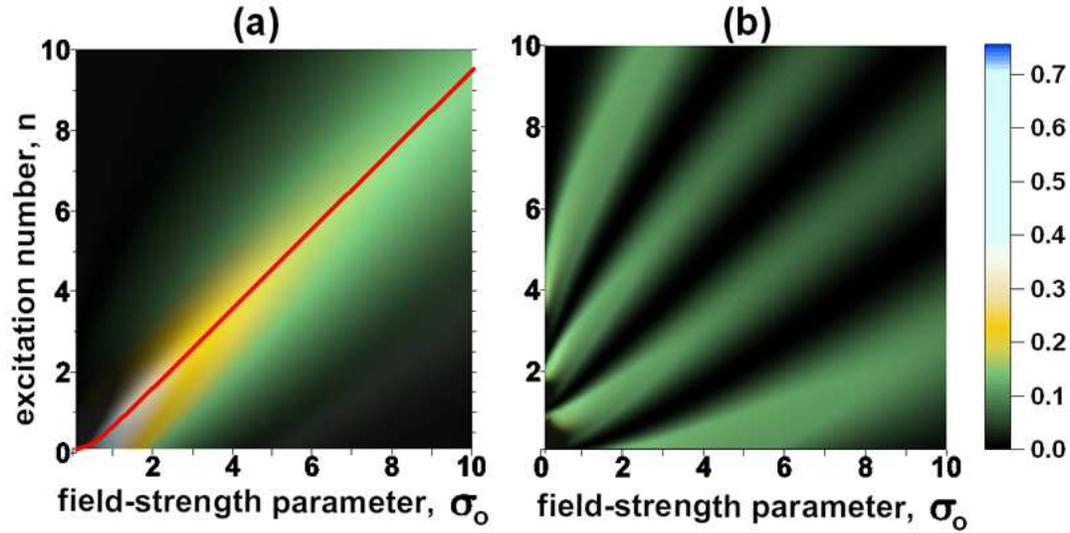}
\caption{The asymptotic distribution $P^{(m\to n)}(t\to\infty)$ of
the two-electron wave packet over the excited states, i.e.\ their
population vs.\ the (analytically continued) quantum number and the
field intensity is illustrated as a 2D intensity plot. Figures (a)
and (b) correspond to the different initial states with $m=0$
(ground state) and $m=3$, respectively. The color scale of the
population probability is common for both figures. The red line in
(a) [determined by eq.~(\ref{NPmax})] denotes the position (the
quantum number) of the maximally populated state for the given
field-intensity parameter $\sigma_o$ (c.f. eq.~(\ref{sigma})).}
\label{fig.PnA}
\end{figure}

Of particular interest are the asymptotic excitation probabilities
$\lim_{t\to +\infty}P^{(m\to n)}(t)$ since they are experimentally
accessible observables. In Fig.~\ref{fig.PnA} the distribution
$P^{(m\to n)}$ is portrayed as a 2D intensity plot vs the
 (analytically continued) excitation quantum number $n$ and the
field-strength parameter $\sigma$. The latter is proportional to the
laser intensity (see the discussion in the next section). Plots (a)
and (b) correspond to the different initial states with $m=0$
(ground state) and $m=3$ accordingly. For $m=0$ the quantum number
of the maximally populated final state $n^*$ is governed by the
equation
\begin{eqnarray}\label{dPdn}
\frac{\diff}{\diff n}P^{(0\to n)}= \frac{\diff}{\diff n}\left(\frac{1}{n!}
\,\,\sigma^n \eulere^{-\sigma}\right)=0,
\end{eqnarray}
which leads to the simple relation
\begin{eqnarray}\label{NPmax}
\sigma=\eulere^{\Lambda(n^*+1)}=
n^*+\frac{1}{2}+{o}\left(10^{-1-\log n^*}\right)\ .
\end{eqnarray}
$\Lambda(n)=\left.\frac{\diff^{n}}{\diff x^{n}}\Gamma(x)\right|_{x=0}$ is the
Euler polygamma function.

\section{Energy absorption}
\label{Sec.EA}
Let us now calculate the energy absorbed from the laser field by the
two-electron system initially prepared in the ground state
$E_\mathrm{abs}\!=\!E(t\!\to\!\infty\!)-E(t\!=\!t_0)$, where $E(t) =
\langle \psi_0(t) | \hat{H}(t) | \psi_0(t)\rangle$ with $\hat{H}(t)$
the total Hamiltonian\footnote{We disregard the possible spontaneous
decay of the excited system.}
\begin{eqnarray}\label{Transp}
\fl\qquad \langle \psi_0(t) | \hat{H}(t) | \psi_0(t)\rangle=
\frac{\epsilon^2\omega^2}{2\mu
c^2}\left(\!\!\left[g^2(t)+f^2(t)\right]\!+\!
2f(t)\frac{A(t)}{\omega}\!+\!\frac{A^2(t)}{\omega^2}\!\right)\!+\frac{\hbar\omega}{2},
\end{eqnarray}
\begin{eqnarray}\label{Eabs}
E_\mathrm{abs}= \lim_{t\to\infty}\,\frac{\epsilon^2\omega^2}{2\mu
c^2}\left[g^2(t)+f^2(t)\right]=
\hbar\omega\lim_{t\to\infty}\sigma(t).
\end{eqnarray}
This expression for the absorbed energy is obtained independently of the initial state, i.e.\ using in eq.~\reff{Transp} the wave function $|\psi_m(t)\rangle$ (starting from an initial state $m$) also leads to the result \reff{Eabs}. The absorbed energy depends on
$\lim_{t\to\infty}\sigma(t)$ only but not on the initial state or the
 population over the final states of the system.
The parameter of central importance thus clearly is
$\sigma(t\to\infty)\equiv\sigma\!_o$, which can be recast into
\begin{eqnarray}\label{sigma}
\sigma_o=\lim_{t\to\infty} \frac{\epsilon^2\omega}{2\mu\hbar
c^2}\left[g^2(t)+f^2(t)\right]= \frac{\epsilon^2\omega}{2\mu\hbar
c^2}\,\lim_{t\to\infty}\left|g(t)+\imagi
f(t)\right|^2=\nonumber\\=\frac{\epsilon^2\omega}{2\mu\hbar c^2}\,
\left|\int_{\!-\infty}^{+\infty}\hspace{-2mm}\diff t'A(t')\eulere^{i\omega
t'}\right|^2=\frac{\epsilon^2\omega}{2\mu\hbar
c^2}\,\,|A(\omega)|^2,
\end{eqnarray}
showing that the spectral power of the incoming laser field {\em
evaluated at the harmonic oscillator frequency} determines the
energy absorption $E_\mathrm{abs}$.

The expression for the absorbed energy \reff{Eabs} can be alternatively derived via the average number
$\Braket{n}$ of absorbed photons calculated using the photon
statistics:
\begin{eqnarray}\label{Eabsn}
E_\mathrm{abs}=\hbar\omega\Braket{n}= \hbar\omega\sum_{n=0}^\infty
n\,\frac{\sigma_o^n}{n!}\,\,\eulere^{-\sigma\!_o}=
\hbar\omega\sigma\!_o.
\end{eqnarray}

The essential fact is that the field strength parameter contains the
cumulative mass and the charge of the CM subsystem with
different powers $\sigma\sim\frac{\epsilon^2}{\mu}$. Therefore, the
observables in the single and in the double-electron excitation
differ from each other. In particular, the energy absorption from
the same laser pulse by a (2e)-system is twice as large as in the one
electron case.

In the generalized many-particle problem the parabolic potential
contains an arbitrary number of electrons $N$. As in the (2e)-case
the $N$-electron CM subhamiltonian accounting for the EM field can be
separated from the field-independent relative motion of the
electrons. Then, all the formulae describing two-electron
transitions are valid for the many-electron case up to the effective
charge and mass. Hence the energy absorption changes linearly with the
number of electrons in the well,
$E^{(Ne)}_\mathrm{abs}/E^{(1e)}_\mathrm{abs}= N$. Given this
property, the number of active electrons in a harmonic system may be
experimentally accessible.

Let us now discuss the two extreme cases of complete survival and full depletion of the initial
state, that is its maximum and minimum survival probability. If
the survival probability at the detection time equals unity, $P{^{(m\to
m)}}(t\to\infty)\to1$, the quantum system does not absorb energy
from the laser field and becomes transparent for the corresponding
pulse. This is strictly realized at the zero value of the
field-strength parameter $\sigma\!_o=0$ in the exponent in
(\ref{Pmn}). Hence, according (\ref{sigma}), the absence of the
oscillator frequency in the spectrum of the vector potential,
$A(\omega)=0$, precludes absorption, [cf.~eq.~(\ref{Eabs})]. Note,
that this statement is correct for any field intensity and not just
in a first order perturbative treatment as for nonparabolic
confinements.

As an example, consider a laser field consisting of two consecutive
Gaussians shifted in time by the interval $2a$:
$A(t)=\sin[\omega_l(t+a)]\eulere^{-\frac{1}{2}\left[(t+a)/t_p\right]^2}\!\!+\,
\sin[\omega_l(t-a)]\eulere^{-\frac{1}{2}\left[(t-a)/t_p\right]^2}$.
Setting e.g., the laser frequency $\omega_l=1$ and the halfwidth of
pulses $t_p=\sqrt{2}/2$, we aim at finding a relation between the
time delay $2a$ and the oscillator frequency $\omega$, for which
$0=A(\omega;a)=-\imagi\sqrt{\pi\slash\!\sqrt{e}}\,\,\cos\omega a\,
\sinh\frac{\omega}{2}\, \eulere^{-{\omega^2}/{4}}$. The
result is $\ \omega a =\pi(k+1/2)\,$ with $k$ an integer. This is,
in fact, a condition for the destructive interference (in time) of
the two waves within the characteristic time-domain $2\pi/\omega$ of
the system. In terms of the excitation picture the explanation is
the following: the electronic wave packet, distributed over the
excited levels by the first pulse, may be assembled back to the
initial state by the second pulse, if the phase difference between
the pulses is chosen properly.

The opposite case amounts to the minimum survival probability, i.e.\ the
maximum depletion of the initial state. On account of
$\eulere^{-\sigma_o}\to 0$ this is the case if the spectral power at
the transition energy (coinciding with the oscillator frequency
$\omega$) is infinitely large. It is equivalent to the presence of an
infinite, monochromatic laser pulse $A(t)=A_0\eulere^{-i\omega_l t}$
with $A(\omega)\sim A_0\,\delta(\omega-\omega_l)$. Such a pulse
resonantly depletes, for instance, the ground state. For a finite
pulse, e.g., $A(t)\!=\!\sin\omega_l t\,\,
\eulere^{-\frac{1}{2}(t/t_p)^2}$ with a power spectrum
$E_\mathrm{abs}\sim|A(\omega)|^2\!=\!2\pi
t_p^2\sinh^2(\omega\omega_lt_p^2)\eulere^{-(\omega^2+\omega_l^2)t_p^2}\neq
0$, there is always nonvanishing absorption. However, only for the
resonance $\omega_l=\omega$ the absorption exists in the limit of
large pulse durations, $t_p\to\infty$.

In addition, if the initial state differs from the ground state, its
full depletion occurs at such values of $\sigma_o$ satisfying
$\sigma_o^m\!\cdot{_2F_0}^{\!}\!\left(-m,-n;-\frac{1}{\sigma_o}\right)=0$.
For example, the well prepared state with $m=1$ will be exhausted
and distributed over other levels by the laser field of the strength
$\sigma\!_o=1$ [cf.~eq.~(\ref{Pmn})]: $P^{(1\to 1)}=
(\sigma\!_o-1)^2\,\, \eulere^{-\sigma\!_o}$. Similar relations can
be obtained for other higher order initial states, e.g.\
$\sigma\!_o=2\pm\sqrt{2}$ for $m=2$,
$\sigma\!_o-3-\alpha-\frac{3}{\alpha}=
2\sigma_o-6+\alpha-\frac{3}{\alpha}\pm
\imagi\sqrt{3}\left(\alpha-\frac{3}{\alpha}\right)=0$,
$\alpha=\left(3+3\imagi\sqrt{2}\right)^{1/3}$ for $m=3$ and so on.

\section{Conclusion} \label{sec5}
In this paper we considered the exactly solvable problem of the
excitation of two interacting electrons, confined in a parabolic
well in the presence of a laser field. The expression for the
time-dependent population of the excited two-electron states was
derived. The physical features of the process, such as the Poisson
statistics of the transition probabilities and the linear dependence
of the energy absorption on the laser power, were analyzed, in
particular, the conditions for the complete survival and full
depletion of the initial state. Our approach can be used for the
testing of approximate theories dealing with few-body multiphoton
excitations.

\section{Acknowledgments}
One of us (O.K.) would like to thank A.\ Voitkiv and N.\ Fominykh for
stimulating discussion. This work was supported by the Deutsche Forschungsgemeinschaft.

%

%
%

\end{document}